\documentstyle[aps,graphicx]{revtex}

\begin{document}
\draft

\title{Measurement of Longitudinal Spin Transfer to $\Lambda$ Hyperons in 
Deep-Inelastic Lepton Scattering}

\author{ 
A.~Airapetian,$^{31}$
N.~Akopov,$^{31}$
M.~Amarian,$^{26,31}$
E.C.~Aschenauer,$^{6}$
H.~Avakian,$^{10}$
R.~Avakian,$^{31}$
A.~Avetissian,$^{31}$
B.~Bains,$^{15}$
C.~Baumgarten,$^{22}$
M.~Beckmann,$^{12}$
S.~Belostotski,$^{25}$
J.E.~Belz,$^{27,28}$
Th.~Benisch,$^{8}$
S.~Bernreuther,$^{8}$
N.~Bianchi,$^{10}$
J.~Blouw,$^{24}$
H.~B\"ottcher,$^{6}$
A.~Borissov,$^{5,14}$
M.~Bouwhuis,$^{15}$
J.~Brack,$^{4}$
S.~Brauksiepe,$^{12}$
B.~Braun,$^{8,22}$
B.~Bray,$^{3}$
S.~Brons,$^{6,28}$
W.~Br\"uckner,$^{14}$
A.~Br\"ull,$^{14,19}$
E.E.W.~Bruins,$^{19}$
H.J.~Bulten,$^{24,30}$
G.P.~Capitani,$^{10}$
P.~Carter,$^{3}$
P.~Chumney,$^{23}$
E.~Cisbani,$^{26}$
G.R.~Court,$^{17}$
P.F.~Dalpiaz,$^{9}$
E.~De~Sanctis,$^{10}$
D.~De~Schepper,$^{19}$
E.~Devitsin,$^{21}$
P.K.A.~de~Witt~Huberts,$^{24}$
P.~Di~Nezza,$^{10}$
M.~D\"uren,$^{8}$
A.~Dvoredsky,$^{3}$
G.~Elbakian,$^{31}$
J.~Ely,$^{4}$
A.~Fantoni,$^{10}$
A.~Fechtchenko,$^{7}$
M.~Ferstl,$^{8}$
K.~Fiedler,$^{8}$
B.W.~Filippone,$^{3}$
H.~Fischer,$^{12}$
B.~Fox,$^{4}$
J.~Franz,$^{12}$
S.~Frullani,$^{26}$
M.-A.~Funk,$^{5}$
Y.~G\"arber,$^{6,8}$
H.~Gao,$^{2,15,19}$
F.~Garibaldi,$^{26}$
G.~Gavrilov,$^{25}$
P.~Geiger,$^{14}$
V.~Gharibyan,$^{31}$
A.~Golendukhin,$^{5,22,31}$
G.~Graw,$^{22}$
O.~Grebeniouk,$^{25}$
P.W.~Green,$^{1,28}$
L.G.~Greeniaus,$^{1,28}$
C.~Grosshauser,$^{8}$
M.~Guidal,$^{24}$
A.~Gute,$^{8}$
V.~Gyurjyan,$^{10}$
J.P.~Haas,$^{23}$
W.~Haeberli,$^{18}$
O.~H\"ausser,$^{27,28,}$\cite{author_note1}
J.-O.~Hansen,$^{2}$
M.~Hartig,$^{28}$
D.~Hasch,$^{6,10}$
F.H.~Heinsius,$^{12}$
R.~Henderson,$^{28}$
M.~Henoch,$^{8}$
R.~Hertenberger,$^{22}$
Y.~Holler,$^{5}$
R.J.~Holt,$^{15}$
W.~Hoprich,$^{14}$
% ...
\medskip\\ \centerline{(The HERMES Collaboration)}\medskip
}

\address{ 
% ...
$^2$Physics Division, Argonne National Laboratory, Argonne, Illinois 60439-4843, USA\\
% ...
$^5$DESY, Deutsches Elektronen Synchrotron, 22603 Hamburg, Germany\\
$^6$DESY Zeuthen, 15738 Zeuthen, Germany\\
% ...
$^8$Physikalisches Institut, Universit\"at Erlangen-N\"urnberg, 91058 Erlangen, Germany\\
% ...
$^{10}$Istituto Nazionale di Fisica Nucleare, Laboratori Nazionali di Frascati, 00044 Frascati, Italy\\
% ...
$^{12}$Fakult\"at f\"ur Physik, Universit\"at Freiburg, 79104 Freiburg, Germany\\
% ...
$^{14}$Max-Planck-Institut f\"ur Kernphysik, 69029 Heidelberg, Germany\\
$^{15}$Department of Physics, University of Illinois, Urbana, Illinois 61801, USA\\
% ...
$^{22}$Sektion Physik, Universit\"at M\"unchen, 85748 Garching, Germany\\
% ...
$^{28}$TRIUMF, Vancouver, British Columbia V6T 2A3, Canada\\
% ...
} 

\date{\today}
\maketitle


\begin{abstract}
Spin transfer in deep-inelastic $\Lambda$ electroproduction has
been studied with the HERMES detector using the 27.6 GeV polarized positron
beam in the HERA storage ring. 
For an average fractional energy transfer $\langle z\rangle =0.45$, the 
longitudinal spin transfer from the virtual photon 
to the $\Lambda$ has been extracted. 
The spin transfer along the $\Lambda$ momentum direction is found to be 
$ 0.11 \pm 0.17 ({\rm stat}) \pm 0.03 ({\rm sys})$; similar values are 
found for other possible choices for the longitudinal spin direction of the 
$\Lambda$. 
This result is the most precise value obtained to date
from deep-inelastic scattering with charged lepton beams, 
and is sensitive to polarized up quark fragmentation to hyperon states.
The experimental result is found to be in general agreement with
various models of the $\Lambda$ spin content,
and is consistent with the assumption of helicity conservation
in the fragmentation process.
\end{abstract}

\medskip

\vspace{1cm}

\centerline{PACS numbers: 13.60.Rj, 13.87.Fh, 13.88.+e, 25.30.Dh}


\twocolumn

Spin-dependent deep-inelastic scattering of charged leptons has
provided precise information on the spin structure
of the nucleon. Several inclusive experiments on polarized 
proton and neutron targets \cite{E142,E143,SMC,E154,g1n,g1p} 
have confirmed the EMC result \cite{EMC}, from which it was 
inferred that the quark spins account for only a small fraction of 
the nucleon spin.  
Additional information has been obtained from semi-inclusive 
polarized deep-inelastic scattering experiments, where 
the correlation between the flavor of the struck quark 
and the type of hadron observed 
in the final state allows the separation of the spin contributions of 
the various quark flavors and of valence and 
sea quarks \cite{SMCsemi,semi}. Those    
measurements indicate that the net contribution 
of the up and down sea quarks to the nucleon spin is small. 
However, considerable uncertainties remain in the contributions 
of strange quarks and gluons.  

It has been proposed 
that one could obtain additional information on the 
polarized quark distributions in the baryons  
of the spin 1/2 octet through the production
of $\Lambda$ hyperons in polarized deep-inelastic lepton scattering 
\cite{BURKARDT93,JAFFE96}.
By measuring the polarization of the $\Lambda$'s that are likely to have 
originated from the struck quark (so-called ``current fragmentation''),  
the longitudinal spin transfer $D_{LL'}^\Lambda$ can be determined.
This quantity is defined as the
fraction of the virtual photon polarization transferred to the $\Lambda$.
In the naive quark parton model (QPM) the spin of the $\Lambda$ is 
entirely due to the strange quark, and the up and down quark 
polarizations are zero.   
On the other hand, assuming SU(3) flavor symmetry, the up, down and strange 
quark distributions (and fragmentation functions) for the $\Lambda$ can 
be related to those in the proton.  
If existing data on hyperon decays and polarized structure functions of 
the nucleon are interpreted in the framework of SU(3) symmetry, the first  
moments of the polarized up and down quark distributions in the 
$\Lambda$ can be estimated to be about -0.2 each \cite{BURKARDT93}. 
If one assumes in addition that quark helicity is conserved 
in the fragmentation process, one obtains
this same negative value for the expected spin transfer 
from a struck up or down quark to the $\Lambda$. 
A measurement of the spin transfer thus has the potential to provide
information on the spin structure of the $\Lambda$ hyperon.

Longitudinal spin transfer in $\Lambda$ production  has previously been studied 
at the Z$^0$ pole at LEP. In the standard model, strange quarks 
(or quarks of charge -1/3 in general)
produced via Z$^0$ decays have an average polarization of -0.91. 
Both the ALEPH \cite{ALEPH} and OPAL collaborations \cite{OPAL} 
have reported a 
measurement of the $\Lambda$ polarization of about -0.3 for $z > 0.3$. 
(Here, $z$ is the fraction of the available energy carried by the $\Lambda$.)
The interpretation of these data is not unique. 
In Ref.~\cite{FLORIAN97}, for example, 
the LEP data have been confronted with three different scenarios,
all of which describe the results 
reasonably well: the naive QPM of the $\Lambda$ spin structure, 
where only the strange quark carries spin    
and contributes to polarized $\Lambda$ production (subsequently 
referred to as Scenario 1), the SU(3) flavor-symmetric model, in which up and 
down quarks also contribute with a negative sign 
(Scenario 2), and a rather extreme hypothesis, in which all three light
quark flavors contribute equally to the $\Lambda$ polarization (Scenario 3).
An alternative approach may be found in the work of \cite{MA00_LEP,BOROS99}
where calculations of the parton distribution functions
in the $\Lambda$ have been performed in various models.
Predictions are then made for the $\Lambda$ spin transfer. 
These predictions display a marked $z$-dependence which is directly related
to the behaviour of the parton distribution functions in the
$\Lambda$ at large $x$. 
In Ref.~\cite{BOROS99}, the LEP data are found to agree well with 
the prediction,
but only if SU(3) symmetry breaking effects are taken into account.
A third approach 
is presented in the LEP publications \cite{ALEPH,OPAL}. Following the 
prescription of \cite{GUSTAFSON93}, the contribution of heavier hyperon
decays to the $\Lambda$ spin transfer was carefully considered.
When a $\Lambda$ is produced from the decay of another hyperon (such as the
$\Sigma^*$), its polarization will reflect that of its parent; 
the resulting spin transfer from the initial `struck' quark to the $\Lambda$
through this channel
will thus be related to the spin structure of the $\Sigma^*$ rather than
the $\Lambda$. In the LEP analyses,
the fractions of $\Lambda$ baryons arising from various decay channels
were estimated using Monte Carlo simulations in the Lund fragmentation model,
and then combined with the naive QPM
values for the quark polarization in the various hyperons.
The resulting prediction was found to agree well with the data, 
with up to 50\% of the spin transfer arising from heavier hyperon decays.
The influence of the heavier hyperons complicates any simple
interpretation of the spin transfer in terms of the $\Lambda$ spin
structure alone; instead, the data must be viewed in the context of models
describing the spin structure of the several hyperons involved.  In
particular, the $z$-dependence of the spin transfer needs to be
considered, since the influence of heavy hyperon decays will diminish as
$z \rightarrow 1$.

In the $e^+ e^-$ experiments at LEP, all three light quark flavors 
contributed significantly to the production of $\Lambda$ hyperons,
with the strange quark playing the dominant role.
By contrast, $\Lambda$ production in deep-inelastic lepton scattering 
is dominated by scattering on up quarks. 
Hence such experiments provide a means to
distinguish between the various models of the $\Lambda$ spin structure,
and to investigate further the degree of 
helicity conservation in the fragmentation process.

The polarization of $\Lambda$ hyperons can be measured via the 
weak decay channel $\Lambda \rightarrow p \pi^-$,
through the angular correlation of the final state:
\begin{equation} 
\frac{dN_{\mathrm{p}}}{d\Omega} 
\propto 1 + \alpha \vec{P}_{\Lambda} \cdot \hat{p}. 
\end{equation} 
Here $\alpha = 0.642 \pm 0.013$ is the asymmetry parameter of the 
parity-violating weak decay, 
$\vec{P}_{\Lambda}$ is the polarization of the $\Lambda$,
and $\hat{p}$ is the unit vector along the proton momentum 
in the rest frame of the $\Lambda$. 
For a longitudinally polarized lepton beam and an unpolarized target, the
$\Lambda$ polarization 
is given in the quark parton model by \cite{JAFFE96,MULDERS96}
\begin{equation}
	\vec{P}_{\Lambda} =  \hat{q} P_B D(y) {
		{\sum_f e_f^2 q^N_f(x,Q^2) G^{\Lambda}_{1,f} (z,Q^2)} \over
		{\sum_f e_f^2 q^N_f(x,Q^2) D^{\Lambda}_{1,f} (z,Q^2)}
	},
	\label{eqn:plam}
\end{equation}
where $P_B$ is the polarization of the charged lepton beam, 
$-Q^2$ is the squared four-momentum transfer of the virtual photon
with energy $\nu$, and $x = Q^2/2M \nu $ is the Bj\o rken scaling variable
(with $M$ denoting the proton mass).
The fractional energy transferred to the nucleon is $y = \nu / E$ 
(where $E$ is the lepton beam energy), 
$z = E_\Lambda/ \nu $ is the energy fraction of the $\Lambda$, and 
$D(y) \approx y (2-y) / (1 + (1-y)^2)$  is the virtual 
photon depolarization factor.
Finally, $q^N_f(x,Q^2)$ is the quark distribution for flavor $f$ in
the nucleon,
$ D^{\Lambda}_{1,f} (z,Q^2)$ is the spin-independent
fragmentation function for $\Lambda$ production from quark flavor $f$, 
$G^{\Lambda}_{1,f} (z,Q^2)$ is the corresponding longitudinal spin-transfer
fragmentation function, and
$e_f$ is the quark charge in units of the elementary charge $e$.
The symbol $\hat{q}$ repesents the unit vector along the virtual 
photon direction. The $\Lambda$ polarization may in general be directed 
along some other axis $\hat{L'}$, 
such as the $\Lambda$ momentum \cite{MULDERS96} or 
the lepton beam momentum \cite{JAFFE96}. However, as $\Lambda$ production 
at the kinematics of the HERMES experiment may be treated as an essentially 
collinear process,  the effects of such 
complexities should be small.  

Following Ref.~\cite{JAFFE96} the component of the longitudinal 
spin transfer to the $\Lambda$ along a longitudinal spin quantization axis 
$L'$ is defined as
\begin{equation}
	D_{L L'}^\Lambda 
	\equiv \frac{\vec{P}_{\Lambda} \cdot \hat{L'}}{P_B D(y)} 
	= \frac{ \sum_{f}e_f^2 q^N_f(x,Q^2)G^{\Lambda}_{1,f}(z,Q^2) }
	       { \sum_{f}e_f^2 q^N_f(x,Q^2)D^{\Lambda}_{1,f}(z,Q^2) },
	\label{eqn:dll}
\end{equation}
where the subscripts $L$ and $L'$ denote the fact that the spin is transferred 
from a polarized photon to a polarized $\Lambda$ and that the two 
longitudinal spin quantization axes may be different.
Due to the charge factor for the up quark,  
the spin transfer in $\Lambda$ electroproduction is dominated by the 
spin transfer from the up quark to the $\Lambda$. 
Moreover, due to isospin symmetry the spin transfer coefficients
from the up and down quarks to the $\Lambda$ are expected to be equal. 
Thus Eq.~(\ref{eqn:dll}) can be approximated by:
\begin{equation}
	D_{L L'}^\Lambda \simeq \frac
		{ G^{\Lambda}_{1,u} (z,Q^2) } 
		{ D^{\Lambda}_{1,u} (z,Q^2) }.
\end{equation}
Consequently, $\Lambda$ electroproduction in the current 
fragmentation region is most sensitive to the 
ratio $G^{\Lambda}_{1,u}  / D^{\Lambda}_{1,u}$
$\sim G^{\Lambda}_{1,d}  / D^{\Lambda}_{1,d}$.
Since the $Q^2$ range of the measurement reported here is small, 
it is assumed that $D_{L L'}^\Lambda$ 
depends only on the energy fraction $z$.  
If the fragmentation process does indeed possess some degree of 
helicity conservation between the struck quark and the final state
hyperon (as supported by the non-zero $\Lambda$ polarization observed at LEP), 
the ratio $G^{\Lambda}_{1,u}  / D^{\Lambda}_{1,u}$
should be related to the polarization 
$\Delta u^\Lambda / u^\Lambda$ of the up quark in the $\Lambda$.
If a significant fraction of the $\Lambda$'s are produced from 
the decays of heavier hyperons, then the 
the ratio $G^{\Lambda}_{1,u}  / D^{\Lambda}_{1,u}$ will be related
instead to a linear combination of the $u$ quark polarizations in the
various hyperons involved.
  
The measurement was carried out by the HERMES experiment at DESY using 
the 27.6 GeV polarized positron beam of the HERA storage ring.
At HERA, the positrons become transversely polarized by the emission of  
synchrotron radiation \cite{sokolov}. 
Longitudinal polarization of the positron beam at the interaction point 
is achieved with spin rotators \cite{BARBER95} 
situated upstream and downstream of the HERMES experiment. 
Equilibrium polarization values in the range of 0.40 to 0.65 are reached 
with a rise time of about 20 minutes.  
The beam polarization is continuously measured using Compton back-scattering 
of circularly polarized laser light. The statistical accuracy of this 
measurement is typically 1\% in 
60 seconds; its systematic uncertainty is 3.4\%, 
dominated by the normalization uncertainty determined 
from the rise-time calibration \cite{BARBER93,MOST97}.
The beam helicity was reversed between the two years of data aquisition.
The data for this analysis are combined from two three-week running periods,  
one in each of 1996 and 1997, which were dedicated to measurements with 
unpolarized targets of hydrogen, deuterium, $^3$He and nitrogen with a typical 
target density of around $1 \cdot 10^{15}$ nucleons/cm${}^2$. 

A detailed description of the HERMES spectrometer
is provided in Ref.~\cite{HERMES}.
The trajectories of the particles are determined in the region in front 
of the spectrometer magnet by a set of two drift chambers, and the momenta 
are determined by matching these to tracks in two sets of drift chambers 
in the back region behind the magnet.  In addition there are 
three proportional chambers inside the magnet to track low momentum 
particles that do not reach the rear section of the spectrometer.
Particle identification is accomplished using a lead glass calorimeter, 
a scintillator hodoscope preceded by two radiation lengths of lead,  
a transition radiation detector, 
and a C$_4$F$_{10}$/N$_2$ (30:70) gas threshold Cerenkov counter.  
Combining the responses of these detectors in a likelihood method 
leads to an average positron identification efficiency of 99\%, with a 
hadron contamination of less than 1\%. In addition, 
the Cerenkov counter is used to distinguish pions from heavier hadrons 
for momenta between 4.5 and 13.5 GeV. 

Semi-inclusive $\Lambda$ events were selected by requiring at least 
three reconstructed tracks: a positron track in coincidence with 
two hadron candidate tracks of opposite charge. Both the track of 
the scattered positron and that of the positive hadron candidate are 
always reconstructed using all drift chambers and all particle identification 
detectors. The negative hadron candidate is allowed to have only 
partial track information.  These partial tracks are reconstructed by the 
drift chambers in the front region and by the wire chambers located 
in the magnet region. In this way it is possible to get momentum and charge 
information from these tracks, though no information from the particle 
identification and drift chambers in the back portion of the spectrometer 
exists. As almost all negative 
particles are pions, particle identification is not essential for these tracks. 
In this analysis, the track resolution at HERMES is dominated by the resolution 
of the drift chambers in front of the magnet.  Thus
the resolution of the partial tracks does not differ significantly from 
that of the full tracks. An invariant mass is reconstructed 
assuming that the positive hadron is a proton and the negative hadron
is a pion.  If more than one positive or negative hadron exists in one
event, all possible pairings of positive and negative hadrons are used. 

\begin{figure}[htb]
  \begin{center}
	\includegraphics[width=0.45\textwidth]{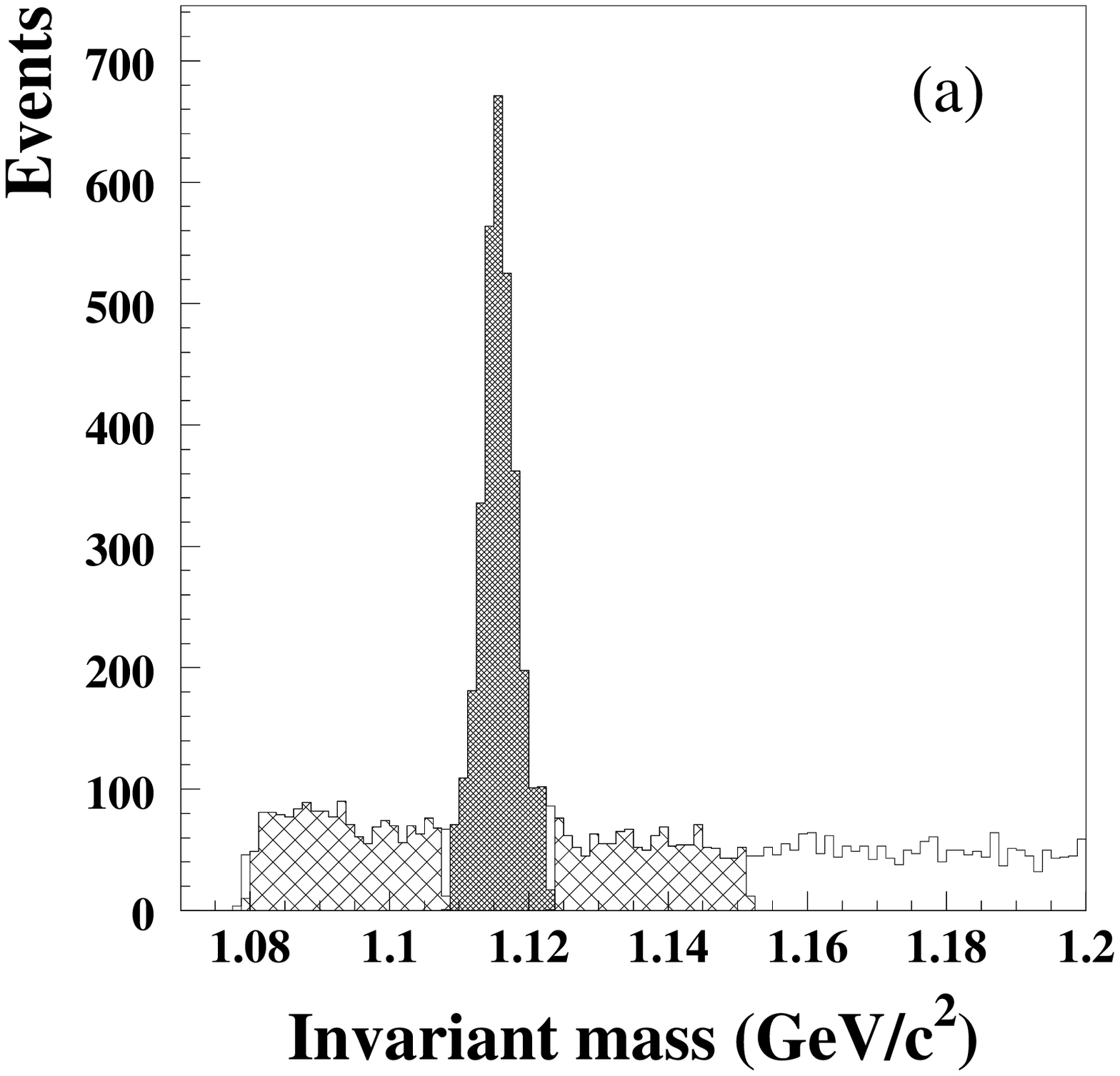}\\[1em]
	\includegraphics[width=0.45\textwidth]{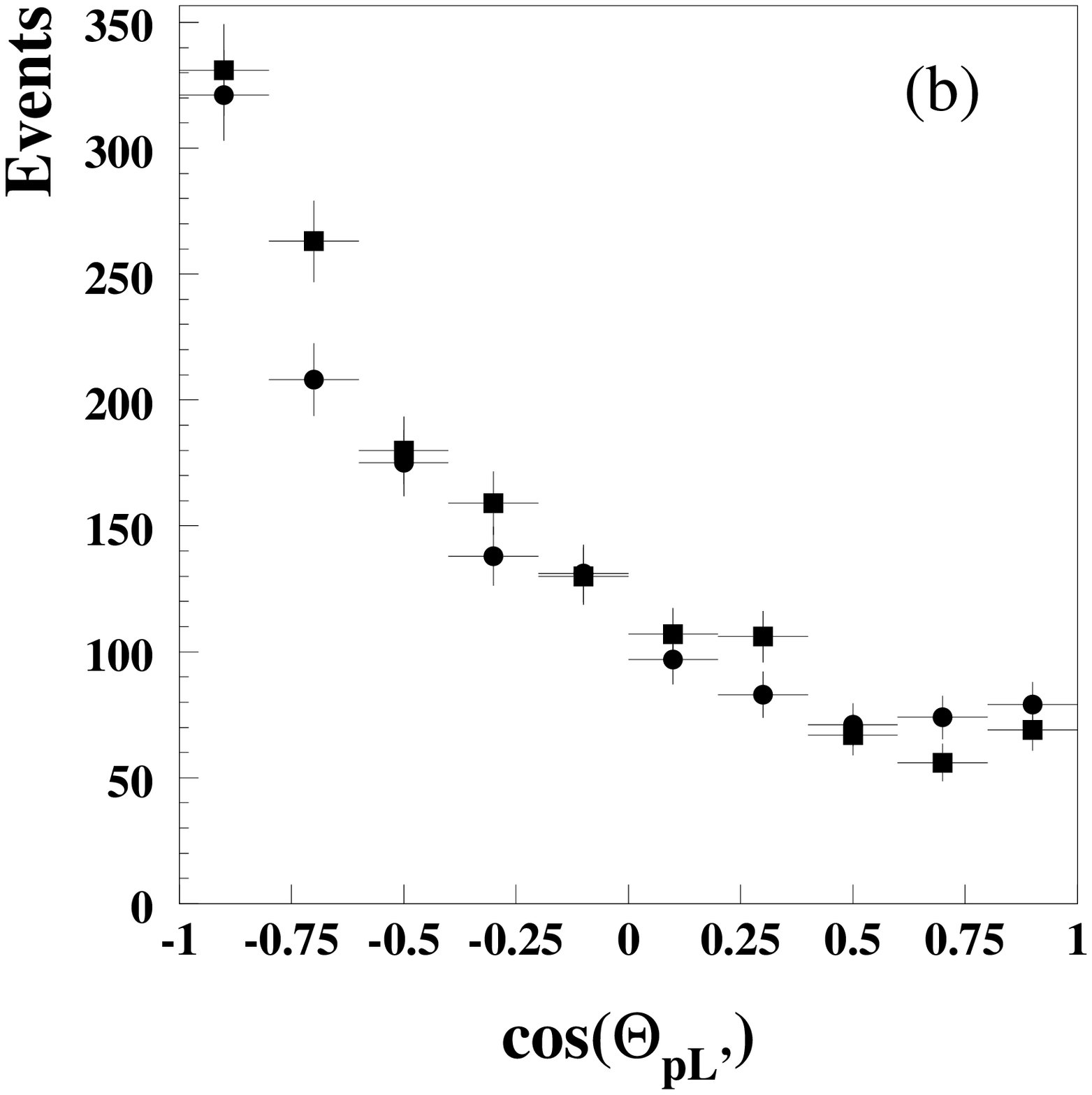}
  \end{center}
  \caption{(a)  Invariant mass spectrum from the reconstruction of candidate
	$\Lambda$ events.  The filled and hatched areas respectively 
	indicate the signal and background regions used in the analysis.
	(b) Spectrum of $\cos{\Theta_{pL'}}$ for the two beam 
	helicities (circles and squares).  The asymmetric appearance of these 
	spectra is almost entirely due to the acceptance of
	the HERMES spectrometer for reconstructing $\Lambda$ decays.}
  \label{fig:one}
\end{figure}

Several requirements were imposed to ensure that
the events are in the  deep-inelastic scattering region and to reduce the
background in the semi-inclusive $\Lambda$ sample: 
$Q^2 >$ 1 GeV$^2$, $W >$ 2 GeV, 
and $y < 0.85$, the latter to avoid a region where radiative 
corrections might be significant.  Here $W$ is defined as the invariant
mass of the photon-nucleon system.  
The calorimeter energy deposited by the scattered
positron was required to be greater than 4 GeV, 
well above the trigger threshold of 3.5 GeV. 
To ensure that the event occurred in the target gas,
the longitudinal vertex position of the positron track was constrained 
to be within the total length of the target cell ($\pm 20$ cm 
from the center of the target). 
The positron interaction vertex and the $\Lambda$ decay vertex were 
required to be separated by more than 10 cm to eliminate background hadrons
originating from the primary vertex. 
Additionally, the distance of closest approach 
between the two hadron tracks was required to be less than 1.5 cm. 
The collinearity, defined as the  
cosine of the angle between the $\Lambda$
momentum (computed from the proton and pion momenta) 
and the $\Lambda$ direction of motion (computed from the vector displacement 
between the positron vertex and the $\Lambda$ decay vertex),  
was required to be above 0.998.
To reduce the large pion contribution to the positive hadron sample, 
the positive hadron was required to have no Cerenkov signal. 
Finally, to ensure that the $\Lambda$ hyperons are primarily from the 
current fragmentation region, 
a positive value of $x_F \approx 2 p_L/W$ was required. 
Here $p_L$ is the momentum component of the $\Lambda$ that is longitudinal 
with respect to the 
virtual photon in the photon-nucleon center-of-mass frame. 
After all these criteria have been implemented, a clean $\Lambda$ signal 
is observed in the invariant mass distribution
(see Fig.~\ref{fig:one}a).
Lambda events have been selected by a cut on the invariant mass distribution: 
 1.109 GeV $< M_{p\pi} <$ 1.123 GeV, resulting in a total number of 
2237 $\Lambda$ events (after background subtraction). 

As the HERMES spectrometer is a forward detector, its acceptance for the 
reconstruction of $\Lambda$ hyperons is limited and strongly depends on  
$\cos\Theta_{pL'}$ (see Fig.~\ref{fig:one}b).
Here $\Theta_{pL'}$ is the angle between the proton 
momentum and the $\Lambda$ spin quantization axis in the rest frame of 
the $\Lambda$. 
To minimize acceptance effects, the spin transfer to the $\Lambda$
has been determined by combining the two data sets measured with opposite 
beam helicities in such a way that the 
luminosity--weighted average beam 
polarization for the selected data sample is zero. 
Using this data sample  and assuming that the spectrometer acceptance 
did not change between the two beam helicity states,
the spin transfer to the $\Lambda$ is determined from 
the forward-backward asymmetry in the angular distributions in 
$\Lambda$ electroproduction \cite{stan,gunar}:  
\begin{equation}
	D_{L L'}^\Lambda = 
	  \frac{1}{\alpha \langle P^2_B \rangle} \cdot 
	  \frac{\sum_{i=1}^{N_\Lambda} P_{B,i} \cos\Theta^i_{pL'}}
               {\sum_{i=1}^{N_\Lambda}D(y_i)\cos^2\Theta^i_{pL'}}.
  \label{eqn:extract}
\end{equation}
The indicated sums are over the $\Lambda$ events, 
and $\langle P^2_B \rangle$ is the luminosity--weighted average of 
the square of the beam polarization. 
The extracted quantity $D_{L L'}$ represents the component of the spin 
transfer coefficient along a chosen quantization axis $L'$, which has been
taken to be parallel to the direction of motion of the $\Lambda$ baryon.
As mentioned earlier, the two similar directions that
have also been considered in this analysis may 
be considered equivalent hypotheses for the true 
direction of the $\Lambda$ polarization, given the collinear nature of the
process at the kinematics of this experiment. 
In addition, 
the derivation of Eq.~(\ref{eqn:extract}) requires that there is
no correlation among the kinematic variables, $i.e.$ between $y$ and 
$\cos\Theta_{pL'}$.  This has been verified for this data set. 

Eq.~(\ref{eqn:extract}) provides a simple method to extract
$D_{L L'}$, without any influence from the spectrometer acceptance, 
from a cross section of the form
\begin{equation}
	\frac{ dN_{p} }{ d\Omega } \propto 
	  1 + \alpha P_B D(y) D_{L L'} \cos\Theta_{pL'}.
	\label{eqn:nomess}
\end{equation}
However, this form of the cross section presupposes that the selected
spin quantization axis $L'$ is indeed the direction of the $\Lambda$
polarization. In general, other components of the polarization may exist.
In this case, the extracted result for $D_{L L'}$ may be contaminated
by interference of certain additional terms in the cross section with 
higher-order terms in the HERMES angular acceptance (see Appendix). 
However, Monte Carlo studies reveal that even if these other components
of the polarization were of the same magnitude as $D_{L L'}$, 
they would contribute to the result presented here
at a level of less than 10\% of the extracted value.

After applying all the requirements described above, 
the longitudinal spin transfer to the $\Lambda$ 
was extracted using Eq.~(\ref{eqn:extract}).  
As no nuclear effects 
were observed within the limited statistics of this measurement,
the data collected on the various targets ($^1$H, $^2$H, $^3$He and $^{14}$N)  
have been added. 
To minimize possible acceptance-induced false asymmetries, 
the data have been corrected for the difference in tracking 
efficiencies between the two years by normalising the number of $\Lambda$    
events to the number of all events 
where two hadrons and a scattered positron were reconstructed.
The  spin transfer $D_{L L'}^\Lambda$ due to 
background events in the selected invariant  mass region has been 
determined from the 
events above and below the $\Lambda$ invariant mass region 
(indicated by the hatched areas in Fig.~\ref{fig:one}a). 
It was found to be consistent with zero  and has been taken into account 
as a dilution. At an average $z$ value of 0.45 the spin transfer 
to the $\Lambda$ is found to be  
$D_{L L'}^\Lambda = 0.11 \pm 0.17$(stat)$\pm 0.03$(sys), 
using the $\Lambda$ momentum as the spin quantization axis $L^{'}$.  
If instead the virtual photon (positron beam) momentum is chosen as 
quantization axis,
the result changes to $0.03 (0.09)$ with the same uncertainties. 
Eqs.~(\ref{eqn:plam}) and (\ref{eqn:dll}) are
based on the assumption that the $\Lambda$ hyperons 
originate from the current fragmentation region. 
Contributions from the target fragmentation 
region are suppressed by the requirement $x_F > 0$ and have been estimated 
by a Monte Carlo simulation to be smaller than 1\%.
The data cover a $z$ range of $0.2 < z < 0.7$,
with $x$ values of $0.02 < x < 0.4$, and with $Q^2$ 
varying between 1 and 10 GeV$^2$. 
The average values of these kinematic variables are
$\langle z \rangle = 0.45$,
$\langle x \rangle = 0.08$,
and $\langle Q^2 \rangle = 2.5$ GeV$^2$.
The systematic uncertainty of the measurement is dominated by the 
uncertainties in detector efficiency differences between the two data sets.  
Possible efficiency differences due to the different kinematic distributions 
of the $\Lambda$ decay 
products and of two reconstructed hadrons in any event 
have been explored and found to be negligible.
Finally, possible false asymmetries induced 
by changes in the detector performance between the two years were investigated 
using both Monte Carlo simulations and samples of hadron pairs 
outside the $\Lambda$ mass peak. 
No significant asymmetries were found by these studies.

\begin{figure}[ht]
  \begin{center}
	\includegraphics[width=0.45\textwidth]{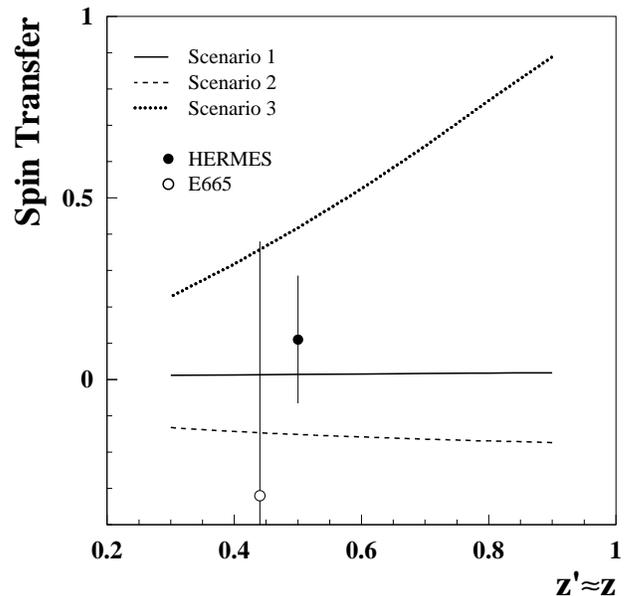}
  \end{center}
  \caption{Spin transfer $D^\Lambda_{LL'}$ as a 
	function of $z^{'} = \frac{E_{\Lambda}}{E_N(1-x)}$ for the spin 
	quantization axis $L'$ along the $\Lambda$ momentum. 
	The error bar represents the quadratic sum of statistical
	and systematic uncertainties.
	The curves correspond to various models for the $\Lambda$ spin 
	structure from Ref.~\protect\cite{FLORIAN97}: 
	the naive QPM (Scenario 1), the SU(3) flavor-symmetric model 
	(Scenario 2), and a model with equal contributions of all light quark 
	flavors to the $\Lambda$ polarization (Scenario 3).}
  \label{fig:two}
\end{figure}

The three models of Ref.~\cite{FLORIAN97} for the $\Lambda$ spin structure 
(mentioned earlier in comparison to the LEP data) have been 
used to predict the $z$ dependence of the spin transfer in $\Lambda$ 
electroproduction. 
In contrast to the LEP data, the DIS measurements are dominated 
by scattering from up quarks and can thus impose different constraints 
on the various $\Lambda$ spin structure scenarios.
Fig.~\ref{fig:two} shows a comparison of the present measurement 
with these predictions. 
Following Ref.~\cite{FLORIAN97} the data are not given at $z=E_{\Lambda}/\nu$  
but at $z^{'} \equiv E_{\Lambda}/(E_N (1-x))$, 
a variable which accounts for the small contamination by target fragmentation.
Here $E_{\Lambda}$ and $E_N$ are the energies of the $\Lambda$ and 
nucleon respectively in the photon-nucleon center of mass system. 
Also shown in the figure is a measurement at a similar $z$ value
from the E665 collaboration \cite{E665}, using DIS with a polarized muon beam. 
The E665 measurement is also similar in its average $Q^2$ value
($\langle Q^2\rangle$ = 1.3 GeV$^2$) but is at much lower $x$
($\langle x\rangle = 0.005$) than the
measurement presented here ($\langle x\rangle = 0.08$).
Further, a recent measurement in $\nu_{\mu}$ charged current interactions
\cite{NOMAD} has shown a $\Lambda$ polarization close to zero in the current
fragmentation region, in agreement with our finding for the $\Lambda$
spin transfer. This measurement is also dominated by $\Lambda$ production
from polarized up quarks.

Fig.~\ref{fig:two} indicates that 
the HERMES measurement appears to favour the 
naive QPM of the $\Lambda$ spin structure (Scenario 1). 
However, as discussed earlier, a significant complication arises from 
the fact that $\Lambda$ hyperons 
may originate from decays of heavier hyperons.
A Monte Carlo estimate shows that only 40-50\% of the $\Lambda$'s
are produced directly; 30-40\% originate from $\Sigma^*(1385)$ decay and
about 20\% are decay products of the $\Sigma^0$.
The up quarks in the $\Sigma^*$ are expected to carry a significant
positive polarization. Polarized up quarks from the target
will thus tend to fragment into $\Sigma^*$ hyperons with a positive
spin transfer coefficient, which is then passed on to the $\Lambda$
polarization through the decay. The $\Sigma^0$ is of lesser influence,
making a smaller contribution of opposite sign to the $\Lambda$ polarization.
The net contribution of $\Sigma$ decays to the $\Lambda$ sample will
thus shift the negative prediction of the SU(3) symmetric model (Scenario 2),
along with that of the naive QPM (Scenario 1), toward positive values.
This effect is illustrated at E665 kinematics in Ref.~\cite{ASHERY99}. 

As pointed out in Ref.~\cite{FLORIAN97}, 
strong contributions from the decays of heavier hyperons provide one possible
cause for the positive spin transfer values of Scenario 3.
Also, large positive values for the spin transfer at high $z$ are
expected in models where the polarization of each quark flavour is expected to
be large in the limit $x \rightarrow 1$ \cite{MA00_HERMES,BOROS99}: 
via the Gribov-Lipatov reciprocity relation \cite{GRIBOV71},
the large quark polarization at $x=1$ produces a large spin transfer
at $z = 1$.
The present HERMES result cannot yet distinguish between these various
models. Additional data will significantly 
improve the precision, and should allow access to higher values
of $z$ where contributions from heavier hyperons are reduced and where
the various models predict markedly different results.

In conclusion, HERMES has measured the longitudinal spin transfer
from the virtual photon to the $\Lambda$ hyperon in deep-inelastic 
electroproduction, finding the value
$D_{LL'} =  0.11 \pm 0.17 ({\rm stat}) \pm 0.03 ({\rm sys})$
at an average fractional energy transfer of $\langle z\rangle = 0.45$.
This result is complementary to measurements from $e^+ e^-$ annihilation, as 
it is uniquely sensitive to the fragmentation of polarized \textit{up} quarks.
The result is in general agreement with calculations based on a variety of
models of the $\Lambda$ spin structure, 
along with the hypothesis of significant helicity conservation in the
fragmentation process (as suggested by earlier data from LEP).
Forthcoming data from HERMES will improve
the precision of the measurement, and help both to explore the $\Lambda$
spin structure and to further test the degree of
helicity conservation in the final state.


\acknowledgments

We gratefully acknowledge the DESY management for its support and the DESY
staff and the staffs of the collaborating institutions.  This work was
supported by the FWO-Flanders, Belgium; 
the Natural Sciences and Engineering Research Council of Canada; 
the INTAS, HCM, and TMR network contributions from the European Community; 
the German Bundesministerium f\"ur Bildung und Forschung;
the Deutscher Akademischer Austauschdienst (DAAD); 
the Italian Istituto Nazionale di Fisica Nucleare (INFN); 
Monbusho, JSPS, and Toary Science Foundation of Japan; 
the Dutch Foundation for Fundamenteel Onderzoek der Materie (FOM); 
the U.K. Particle Physics and Astronomy Research Council; 
and the U.S. Department of Energy and National Science Foundation.


\appendix
\section*{}

As described above, the procedure used to extract the longitudinal
spin transfer coefficient $D_{L L'}$ [Eq.~(\ref{eqn:extract})]
is based on the assumption that the selected spin quantization axis 
$L'$ is indeed the direction of the $\Lambda$ polarization. 
However, if other components of the polarization exist, the extracted 
result for $D_{L L'}$ may be contaminated via interference 
of certain additional terms in the cross section 
with higher-order terms in the HERMES angular acceptance.

Let us introduce 3 perpendicular axes in the $\Lambda$ center of mass
frame, defined by two chosen unit vectors $\hat{J}$ and $\hat{T}$: 
	$\hat{e}_1 \equiv \hat{J}$, 
	$\hat{e}_2 \equiv \hat{J}\times\hat{T}/|\hat{J}\times\hat{T}|$,
	$\hat{e}_3 \equiv \hat{e}_1\times\hat{e}_2$. 
Further, let the symbol $D_{Li}$ refer 
to the probability for spin transfer from a
longitudinally polarized virtual photon to a $\Lambda$ baryon with
polarization along the axis $i$; the symbol $D_{Ui}$ denotes the probability
for $\Lambda$ polarization along the axis $i$ given an unpolarized beam.
We take the vector $\hat{J}$ to represent our direction of interest for
longitudinal spin transfer to the $\Lambda$, namely the direction of the
virtual photon. The quantity $D_{L1}$ is thus identical to the quantity 
$D_{LL'}^\Lambda$ defined previously [Eq.~(\ref{eqn:dll})].
The vector $\hat{T}$ may be either of the other two vectors available: the 
electron beam direction or the momentum of the final state $\Lambda$.
The second axis $\hat{e}_2$ thus represents the direction normal to 
the production plane.
The number of interfering terms is greatly restricted by applying
parity and rotational invariance to a general angular decomposition
of the cross section, and by the fact that the HERMES spectrometer is
symmetric in the vertical coordinate. Finally one is left with only
two terms:
\begin{equation}
	\alpha D_{U2} \cos\Theta_2 \cdot (P_B \sin(n\Phi))
  \label{eqn:interfone}
\end{equation}
and  
\begin{equation}
	\alpha P_B D(y) D_{L3} \cos\Theta_3 \cdot (1 + C_n \cos(n\Phi) ).
  \label{eqn:interftwo}
\end{equation}
In obtaining these expressions, it is important to note that the axis
$\hat{e}_2$, which represents the direction normal to the reaction plane, 
transforms as a pseudo-vector, while $\hat{e}_1$ and
$\hat{e}_3$ transform as vectors.
The first contribution [Eq.~(\ref{eqn:interfone})] depends entirely 
on a non-zero $P_B \sin(n\Phi)$ azimuthal moment in $\Lambda$ production, 
where $\Phi$ denotes the angle between the
$\Lambda$ and the electron scattering plane, around the $\vec{q}$ vector.
Such moments have been measured to be small in pion production. 
In addition, they are coupled here with a transverse $\Lambda$ polarization
and can only appear in the cross section at higher twist 
(i.e. they are suppressed at order $p_T/Q$).
The second term [Eq.~(\ref{eqn:interftwo})] corresponds to the 
other component of $\Lambda$ spin transfer in the production plane, 
and could contribute if the choosen spin quantization axis differs 
dramatically from the true $\Lambda$ polarization direction.
Monte Carlo studies reveal that even if either of the coefficients
$D_{U2}$ or $D_{L3}$ were of the same magnitude as $D_{L1}$, 
they would contribute to the extracted component
at a level of less than 10\% of its value.


\newcommand{\PRL}[3]{Phys. Rev. Lett \textbf{#1}, #3 (#2)}
\newcommand{\PRD}[3]{Phys. Rev. D \textbf{#1}, #3 (#2)}
\newcommand{\PLB}[3]{Phys. Lett. B \textbf{#1}, #3 (#2)}
\newcommand{\NPB}[3]{Nucl. Phys. \textbf{B#1}, #3 (#2)}
\newcommand{\EPJC}[3]{Eur. Phys. J. C \textbf{#1}, #3 (#2)}
\newcommand{\NIMA}[3]{Nucl. Instrum. Methods A \textbf{#1}, #3 (#2)}

\newcommand{\JOURNAL}[4]{#1 \textbf{#2}, #4 (#3)}

\vspace*{0.25cm}
\noindent
* References \cite{stan,gunar}
are available as HERMES internal notes,
and can be accessed on the HERMES web pages:\\[-1.8em]
\begin{center} http://hermes.desy.de/notes/ \end{center}

\end{document}